\documentclass[12pt,a4paper]{article}
\usepackage{graphicx}
\usepackage{subeqnarray}
\begin{document}
%
%
%
%
\newenvironment{lefteqnarray}{\arraycolsep=0pt\begin{eqnarray}}
{\end{eqnarray}\protect\aftergroup\ignorespaces}
\newenvironment{lefteqnarray*}{\arraycolsep=0pt\begin{eqnarray*}}
{\end{eqnarray*}\protect\aftergroup\ignorespaces}
\newenvironment{leftsubeqnarray}{\arraycolsep=0pt\begin{subeqnarray}}
{\end{subeqnarray}\protect\aftergroup\ignorespaces}
\newcommand{\diff}{{\rm\,d}}
\newcommand{\pprime}{{\prime\prime}}
\newcommand{\szeta}{\mskip 3mu /\mskip-10mu \zeta}
\newcommand{\FC}{\mskip 0mu {\rm F}\mskip-10mu{\rm C}}
\newcommand{\appleq}{\stackrel{<}{\sim}}
\newcommand{\appgeq}{\stackrel{>}{\sim}}
\newcommand{\Int}{\mathop{\rm Int}\nolimits}
\newcommand{\Nint}{\mathop{\rm Nint}\nolimits}
\newcommand{\arcsinh}{\mathop{\rm arcsinh}\nolimits}
\newcommand{\range}{{\rm -}}
\newcommand{\displayfrac}[2]{\frac{\displaystyle #1}{\displaystyle #2}}
%
%
\title{Gravitation, holographic principle, and extra dimensions}

\author{
 {R.~Caimmi}\footnote{
{\it Physics and Astronomy Department, Padua University,
Vicolo Osservatorio 3/2, I-35122 Padova, Italy. Affiliated up to September
30th 2014. Current status: Studioso Senior. Current position: in retirement
due to age limits.}\hspace{50mm}
email: roberto.caimmi@unipd.it~~~
fax: 39-049-8278212}
\phantom{agga}}

\maketitle
\begin{quotation}
\section*{}
\begin{Large}
\begin{center}

 Abstract

\end{center}
\end{Large}
\begin{small}

\noindent\noindent
Within the context of Newton's theory of gravitation, restricted to point-like
test particles and central bodies, stable circular orbits in ordinary space
are related to stable circular paths on a massless, unmovable, undeformable
vortex-like surface, under the action of a tidal gravitational field along the
symmetry axis.  An interpretation is made in the light of a holographic
principle, in the sense that motions in ordinary space are connected with
motions on a selected surface and vice versa.   Then ordinary space is
conceived as a 3-hypersurface bounding a $n$-hypervolume where gravitation
takes origin, within a $n$-hyperspace.   The extension of the holographic
principle to extra dimensions implies the existence of a minimum distance
where test particles may still be considered as distinct from the central
body.   Below that threshold, it is inferred test particles lose theirs
individuality and ``glue'' to the central body via unification of the four
known interactions and, in addition, (i) particles can no longer be conceived
as point-like but e.g., strings or membranes, and (ii) quantum effects are
dominant and matter turns back to a pre-big bang state.   A more detailed
formulation including noncircular motions within the context of general
relativity, together with further knowledge on neutron stars, quark stars and
black holes, would provide further insight on the formulation of quantum
gravity.
\noindent

{\it keywords -
Newtonian theory of gravitation - holographic principle - extra dimensions.}
\end{small}
\end{quotation}

\section{Introduction} \label{intro}

According to current QCDM cosmologies, the universe as a whole is
increasingly expanding e.g., [1] which cannot be true for subsystems.  For
instance, massive
(exceeding about 20 solar masses) stars at the end of evolution are no
longer pressure-balanced and collapse into a black hole e.g., [2] where matter
may safely be conceived as turned into a pre-big bang state, implying
unification of the four known interactions as predicted by supersymmetric
theories e.g., [3].   In the light of general relativity, the final
configuration is
point-like but the occurrence of quantum effects could yield strings or
membranes, leaving the gravitational field outside the instability region
(where test particles cannot remain in stable equilibrium)
spherical-symmetric.   Deeper investigation on neutron stars and quark stars
(if their existence is proved), which are celestial bodies most similar to
black holes, could provide further insight on the formulation of quantum
gravity.

A simplified description can be made within the framework of Newton's theory
of gravitation, where spherical-symmetric matter distributions can be
conceived as point-like outside their boundaries.   The concept of space
curvature may be introduced considering a vortex-like surface under the action
of a tidal gravitational field along the symmetry axis, where the distance of
a surface point from the
symmetry axis relates to the distance of a test particle from a central body.
The current paper aims to answer two specific questions restricted, for
simplicity, to circular motions.

First, could stable circular orbits (radius, $R$) described by a test particle
(mass, $m$) around a central body (mass, $M$) be related to stable circular
paths described by a test particle (mass, $m$) on a massless, unmovable,
undeformable vortex-like surface, subjected to a constant gravitational force,
$-GMm/R^2$, directed downwards along the symmetry axis?   If yes, a principle
of corresponding states could be expressed as a holographic
principle in the sense that motions in ordinary space are connected with
motions on a selected surface and vice versa.

Second, could the above mentioned holographic principle be extended to
ordinary space, conceived as a 3-hypersurface, where test particles are
subjected to a constant
gravitational force from extra dimensions?   If yes, could quantum effects and
unification of the four known interactions be interpreted in this context?

The first and second question raised are dealt with in Section \ref{que1} and
\ref{que2}, respectively.   The discussion is drawn in Section \ref{disc}. The
conclusion is shown in Section \ref{conc}.   A special vortex-like
surface and an analogy for the invariance of light velocity are presented in
Appendix \ref{a:guex} and \ref{a:pilv}, respectively.

\section{A holographic principle}
\label{que1}

Let a test particle of mass, $m$, move along a stable circular orbit of
radius, $R$, around a central body of mass, $M$.   Let both the test particle
and the central body be conceived as point-like.   The balance of
gravitational and centrifugal force at a generic point of the orbit, after
little algebra reads:
\begin{lefteqnarray}
\label{eq:eqor}
&& V^2=\frac{GM}R~~;
\end{lefteqnarray}
where $G$ is the constant of gravitation and $V$ the circular velocity.

\begin{figure*}[t]  
\begin{center}      
\includegraphics[scale=0.8]{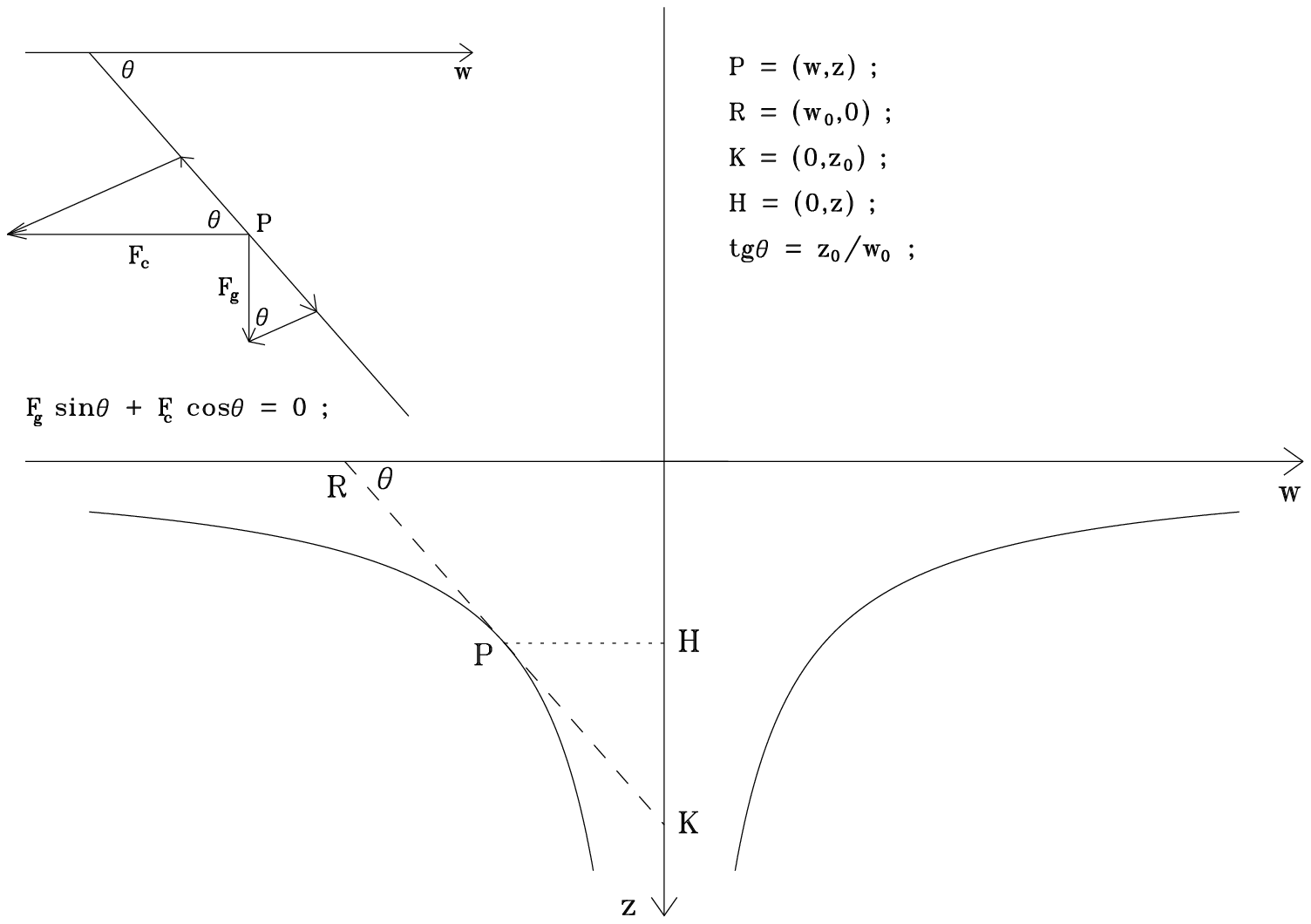}                      
\caption[ddbb]{
A vortex-like, massless, unmovable, undeformable surface subjected to a tidal
gravitational field along the symmetry axis, implying constant attraction
directed downwards.  The axis, $w$, $w^2=x^2+y^2$, lies on the horizontal
plane.   The axis, $z$, coincides with the symmetry axis i.e. the vertical
direction, related to free fall.  The angle, $\theta$, is formed by the axis,
$w$, and the tangent to the surface, lying on a meridional plane and crossing
a selected point, ${\sf P}$.   The tangential component of the gravitational
attraction, $F_g$, and the centrifugal repulsion, $F_c$, related to a test
particle on ${\sf P}$ moving along a stable circular path, $(w,z)$, are
$F_g\sin\theta$ and $F_c\cos\theta$, respectively.   The angle, $\theta$, can
be inferred from the equation of the surface, $z=f(w)$, where
$\lim_{w\to\mp\infty}\diff f/\diff w=0^\pm$ and
$\lim_{w\to0^\mp}\diff f/\diff w=\pm\infty$.   The slope of the tangent is
positive in the third quadrant and negative in the fourth quadrant, as the $z$
axis is oriented downwards.   The curves plotted are branches of equilateral
hyperbola, $wz=\mp10$.  The coordinates of interest and the angle, $\theta$,
are specified on the upper right.   The force balance on a test particle
on the point, ${\sf P}$, is shown on the upper left.}
\label{f:vort}     
\end{center}       
\end{figure*}

Let a vortex-like, massless, unmovable, undeformable surface (in short,
surface) be subjected to a tidal gravitational field along the
symmetry axis, implying constant attraction directed downwards, as depicted
in Fig.\,\ref{f:vort}.
Let the surface be defined as $z=f(w)$, $w^2=x^2+y^2$, with the following
boundary conditions:
\begin{lefteqnarray}
\label{eq:boco}
&& \lim_{w\to\mp\infty}\frac{\diff f}{\diff w}=0^\pm~~;\qquad
\lim_{w\to0^\mp}\frac{\diff f}{\diff w}=\pm\infty~~;
\end{lefteqnarray}
where the vertical axis is directed downwards.   Let ${\sf P}\equiv(w,z)$ be a
generic point on the surface; $w_0$, $z_0$, the intercepts of a tangent to the
surface, lying on a meridional plane and crossing ${\sf P}$, and $\theta$ the
angle defined by the tangent and the axis, $w$, see Fig.\,\ref{f:vort}.

The equation of the tangent is:
\begin{lefteqnarray}
\label{eq:tang}
&& w_0z+z_0w=w_0z_0~~; \\
\label{eq:slta}
&& \tan\theta=-\frac{z_0}{w_0}=-\frac{z_0-z}w~~;\qquad -w\tan\theta>0~~;
\end{lefteqnarray}
where $\theta\to0^\pm$ implies $w\to\mp\infty$, $z\to0^+$, and
$\theta\to(\pi/2)^\mp$ implies $w\to0^\mp$, $z\to+\infty$.   The evaluation of
$\theta$ needs the knowledge of the surface equation.

With regard to a test particle on a selected point, ${\sf P}$, of the surface,
the gravitational force, $F_g$, can be splitted into a normal component,
$F_g\cos\theta$, balanced by the reaction due to surface unmovability and
undeformability, and a tangential component, $F_g\sin\theta$, directed
downwards i.e. towards increasing $z$, see Fig.\,\ref{f:vort}.  Similarly,
the centrifugal force, $F_c$, can be splitted into a normal component,
$F_c\sin\theta$, balanced by the reaction due to surface unmovability and
undeformability, and a tangential component, $F_c\cos\theta$, directed
upwards i.e. towards increasing $\vert w\vert$, see Fig.\,\ref{f:vort}.

Motion along a stable circular path, $\vert w\vert=$ const, $z=$ const,
implies
$F_g\sin\theta+F_c\cos\theta=0$.   The choice, $F_g=GMm/R^2$, translates into:
\begin{lefteqnarray}
\label{eq:eqvo}
&& v^2=-\frac{GM}R\frac wR\tan\theta=\frac{GM}R\frac{z_0-z}R~~;
\end{lefteqnarray}
where $v$ is the circular velocity.

The combination of Eqs.\,(\ref{eq:eqor}) and (\ref{eq:eqvo}) yields:
\begin{lefteqnarray}
\label{eq:eqVv}
&& \frac{v^2}{V^2}=-\frac wR\tan\theta=\frac{z_0-z}R~~;
\end{lefteqnarray}
and the substitution of Eq.\,(\ref{eq:slta}) into (\ref{eq:eqVv}) produces:
\begin{lefteqnarray}
\label{eq:wVv}
&& w=\frac{v^2}{V^2}\frac R{z_0}w_0~~;
\end{lefteqnarray}
where the intercepts of the equation of the tangent, $w_0$ and $z_0$, see
Fig.\,\ref{f:vort}, can be determined using the surface equation, $z=f(w)$.
In the special case where the meridional section of the surface shows branches
of equilateral
hyperbola, one finds $w_0/z_0=w/z$ and Eq.\,(\ref{eq:wVv}) reduces to:
\begin{lefteqnarray}
\label{eq:zRvV}
&& \frac zR=\frac{v^2}{V^2}~~;
\end{lefteqnarray}
for a formal demonstration, an interested reader is addressed to Appendix
\ref{a:guex}.

The above results may be summarized into a single statement.

\begin{trivlist}%
\item[\hspace\labelsep{\bf Holographic principle.~}]
Given a central body of mass, $M$, and a vortex-like, unmovable, undeformable
surface
subjected to a tidal gravitational field along the symmetry axis, implying
constant attraction directed downwards, any test particle moving on a stable
circular orbit of radius, $R$, and velocity, $V$, has a counterpart on the
vortex-like surface, moving along a stable circular path of radius,
$\vert w\vert$, and velocity, $v$, around the symmetry axis, where
$w=(v/V)^2(R/z_0)w_0$.
\end{trivlist}%

In general, the above statement should be conceived as a principle of
corresponding states:
the term ``holographic'' has to be intended in the sense that  motions in
ordinary space are connected with motions on a selected surface and vice
versa.

The above considerations could, in principle, be extended to general
relativity but, for simplicity, are restricted to classical mechanics,
which implies velocities up to infinity.   If light velocity is
assumed as an empirical upper limit, $v\le c$, stable orbits cannot occur
below a critical radius, $R_c$, expressed via Eq.\,(\ref{eq:eqor}) as:
\begin{lefteqnarray}
\label{eq:Rc}
&& R_c=\frac{GM}{c^2}~~;
\end{lefteqnarray}
according to Newton's theory of gravitation.   Its counterpart in general
relativity, using Schwarzschild's metric, reads e.g., [4]:
\begin{lefteqnarray}
\label{eq:RcS}
&& R_c=\frac{3GM}{c^2}~~;
\end{lefteqnarray}
which is three times larger.

Similarly, stable circular paths on a vortex-like surface, centered on the
symmetry axis, cannot occur below a critical radius, $\vert w_c\vert$,
expressed via Eqs.\,(\ref{eq:wVv}) and (\ref{eq:Rc}) as:
\begin{lefteqnarray}
\label{eq:wc}
&& w_c=R_c\frac{w_0}{z_0}=\frac{GM}{c^2}\frac{w_0}{z_0}~~;
\end{lefteqnarray}
where $v=V=c$.

\section{Gravitation and extra dimensions}
\label{que2}

Let ordinary space be conceived as a 3-hypersurface bounding a
$n$-hypervolume, totally $(n=4)$ or partially $(n>4)$.   Let the source of the
gravitational field lie within the hypervolume and let the hypersurface be
curved in presence of matter via gravitational interaction.
To this respect, the following analogy can help. 

With regard to a planet surrounded by a liquid shell, a global ocean
say, the source of the gravitational field within the hypervolume is
represented by the whole planet and matter within the hypersurface by the
water surface, respectively.   In absence of matter (empty ordinary space),
the water surface is totally flat.   In presence of negligible amount of
matter, the water surface can be considered flat to a good extent.   In
presence of considerable amount of matter, the water surface is curved by
gravitational interaction.  A test
particle on the curved surface is subjected to the tangential component of the
gravitational force, as the normal component is balanced by the surface
tension.   In principle, the curvature angle with respect to the horizontal
plane ranges from $\theta=0$ (vortex top) to $\theta=\mp\pi/2$ (vortex
bottom).

Keeping in mind the above mentioned analogy, let the whole range of curvature
be represented as an equilateral hyperbolic funnell i.e. a revolution figure
related to the rotation of the bottom branch of an equilateral hyperbola
around the axis, $z$, as depicted in Fig.\,\ref{f:vort}, where the axis, $w$,
$w^2=x^2+y^2$, lies on the horizontal plane and the axis, $z$, is directed
along the free-fall direction within the hyperspace.    Accordingly, the
equation of the funnell, via Eq.\,(\ref{eq:eqhy}) in Appendix \ref{a:guex},
reads:
\begin{lefteqnarray}
\label{eq:eqfu}
&& (x^2+y^2)z^2=k^2~~;\qquad z>0~~;
\end{lefteqnarray}
due to the symmetry with respect to the rotation axis, $z$.

Let a test particle of mass, $m$, move on a stable circular orbit of radius,
$R$, around a central body of mass, $M$, both conceived as point-like for
simplicity.   Let $F_g$, $F_G$; $F_c$, $F_C$; denote gravitational and
centrifugal forces acting on the test particle within the hypersurface
(ordinary space) and hypervolume (hyperspace), respectively, which implies the
following relations:
\begin{lefteqnarray}
\label{eq:FGC}
&& F_G\sin\theta+F_C\cos\theta=0~~; \\
\label{eq:FG}
&& F_g=F_G\sin\theta=-\frac{GMm}{R^2}~~; \\
\label{eq:FC}
&& F_c=F_C\cos\theta=\frac{mV^2}R~~;
\end{lefteqnarray}
where $G$ is the constant of gravitation and $V$ the orbital velocity.
The combination of Eqs.\,(\ref{eq:FGC})-(\ref{eq:FC}) yields
Eq.\,(\ref{eq:eqor}) and, in particular, Eq.\,(\ref{eq:Rc}) assuming light
velocity as an empirical upper limit.

The gravitational force within the hypervolume via Eq.\,(\ref{eq:FG}) reads:
\begin{lefteqnarray}
\label{eq:FGg}
&& F_G=-\frac{GMm}{R^2\sin\theta}~~;
\end{lefteqnarray}
where $R^2\sin\theta=$ const according to the assumption of hypersurface where
test particles are subjected to a constant gravitational force, $F_G$, from
extra dimensions along the symmetry axis.   Then $\theta\to0$ implies
$R\to+\infty$ and, in addition,
$R^2\sin\theta=R_0^2$, where $R_0$ relates to
$\theta=\pi/2$.   In general, the following relation holds:
\begin{lefteqnarray}
\label{eq:teta}
&& \theta=\arcsin\left(\frac{R_0}R\right)^2~~;
\end{lefteqnarray}
in terms of the orbit radius, $R$.   On the other hand,
$R_0\to0$ can be inferred from general relativity e.g., [5], and a
nonzero $R_0$ can be interpreted as due to quantum effects, which act
to ``close'' the bottom of the funnell.

\begin{figure*}[t]  
\begin{center}      
\includegraphics[scale=0.8]{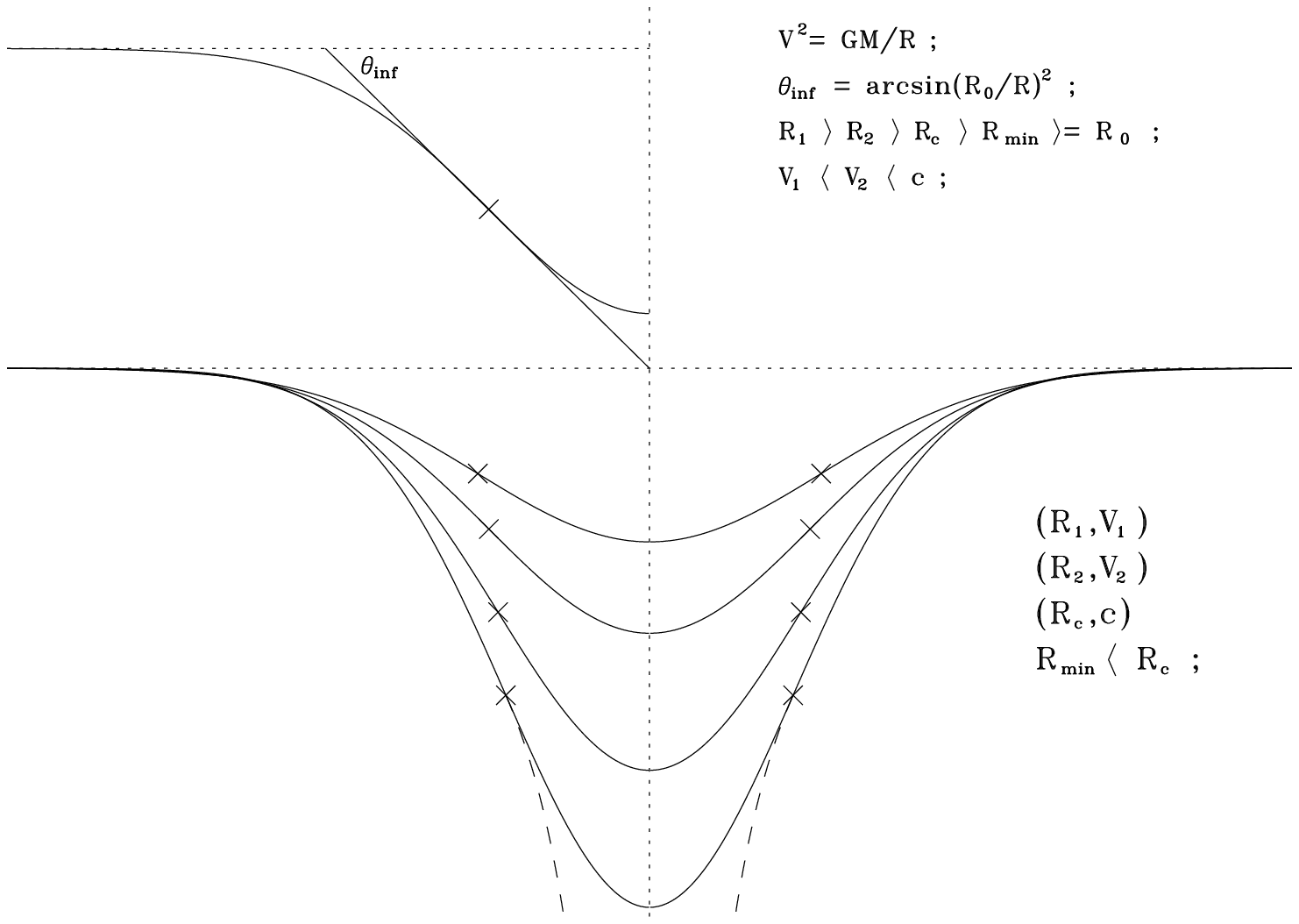}                      
\caption[ddbb]{Qualitative representation of curvature effects induced by a
finite matter distribution, spherical-symmetric for simplicity, defined by a
total mass, $M$, within a finite radius, $R$.   Curvature effects are larger
on the inflexion point of a meridional section of the funnell.   A test
particle in a stable circular orbit on the surface of the matter distribution
has velocity, $V$, inversely proportional to the radius, $R$, until $V=c$ for
$R=R_c$.   Lower radii, $R<R_c$, would imply a vortex-like distortion (dashed)
in absence of quantum effects, which act to ``close'' the bottom of the
funnell up to $R=R_{\rm min}\ge R_0$, where the angle between the tangent on
the inflexion
point and the horizontal axis, with regard to a meridional section (upper
left), is $\theta_{\rm inf}=\theta_{\rm max}\le\pi/2$.   The meridional
section of the funnel has been arbitrarily plotted as an upturned Gaussian,
$z=-a\exp(-bw^2)$, where $(a,b)=(3.8,0.07),(5.8,0.08),(8.8,0.09),(9.8,0.10)$,
from top to bottom, and the
inflexion points are marked by saltires.   Dashed curves are branches of
equilateral hyperbola passing through the inflexion points of the bottom
curve.   See text for further details.}
\label{f:cure}     
\end{center}       
\end{figure*}

Let a finite matter distribution, spherical-symmetric for simplicity, be
defined by a total mass, $M$, within a finite radius, $R$.   Accordingly,
curvature effects are directly proportional to the angle, $\theta$, between
the horizontal axis and the tangent on a selected point, with regard to a
meridional section of the funnell, as depicted in Fig.\,\ref{f:cure}.
Curvature effects are larger on the surface, which can be related to the
angle, $\theta=\theta_{\rm inf}$, involving the inflexion point of related
meridional
section.   Within the matter distribution, curvature effects are reduced due
to Newton's theorem e.g., [6] and removed at the centre, where no gravitational
attraction takes place as (in the current interpretation) the gravitational
force, $F_G$, has null tangential component.

The velocity of a test particle in a stable circular orbit, on the surface of
the matter distribution, is increasing for decreasing $R$ until $V=c$ for
$R=R_c$.   Lower radii, $R<R_c$, would imply a vortex-like distorsion but the
bottom of the funnel is expected to be ``closed'' by quantum effects behind a
threshold, $R=R_{\rm min}\ge R_0$, related to
$\theta=\theta_{\rm max}\le\pi/2$.
For $R<R_{\rm min}$, $\theta<\theta_{\rm max}$, up to $R=0$, $\theta=0$, where
no gravitational attraction takes place, as expected.   Accordingly,
$\theta_c\le\theta_{\rm max}\le\pi/2$, which implies
$R_c\ge R_{\rm min}\ge R_0$, via Eq.\,(\ref{eq:teta}).   If, in particular,
$R_c=(3/2)R_g$, $R_{\rm min}=R_0=R_g$, where $R_g=2GM/c^2$, with regard to
Schwarzschild black holes in general relativity e.g., [7] (Chap.\,XI, \S97),
then Eq.\,(\ref{eq:teta})
reduces to $\theta_c=\arcsin(R_{\rm min}/R_c)^2=\arcsin(4/9)\approx0.1466\pi$.

In summary, stable circular orbits of test particles around a central body
can occur for $R\ge R_c$, implying $V\le c$, while the motion of test
particles is unstable for $R<R_c$ up to $R=R_{\rm min}$.   Beyond that
threshold, quantum effects are expected to be dominant and test particles
are ``glued'' to the central body.

\section{Discussion}
\label{disc}

According to the holographic principle inferred in Section \ref{que1}, stable
circular orbits of a test particle of mass, $m$, around a central body of
mass, $M$, can be related to stable circular paths of a test particle of mass,
$m$, on a vortex-like, massless, unmovable, undeformable surface, subjected
to a tidal gravitational field along the symmetry axis, implying constant
attraction directed downwards.   In other words, the gravitational action on
the ordinary space can be related to the gravitational action on an
axisymmetric surface.   By analogy, the gravitational action on a
$n$-hyperspace can be related to the gravitational action on an axisymmetric
3-hypersurface, coinciding with the ordinary space.   For reasons of
simplicity, considerations have been restricted to circular motions within
Newton's theory of gravitation, but it can safely be expected an extension to
generic motions within the theory of general relativity.

If gravitational attraction takes origin from a $n$-hypervolume, the origin of
(ordinary) space curvature is no longer due to matter in itself, but to
gravitation from extra dimensions.   Then matter appears to be entirely
passive, in the sense that it can no longer be thought of as source of
gravitational field.   In this view, extra dimensions extend as well as
ordinary ones, contrary to assumptions of superstring theories, where extra
dimensions are highly compressed e.g., [3].

The main consequence of the above mentioned analogy is the occurrence of
quantum effects, in the sense that a test particle can approach a central body
(both considered as point-like) up to a threshold, $R=R_{\rm min}$, preserving
its intrinsic properties.   A physical interpretation could be the following.
First, for sufficiently short distances, masses can no longer be conceived as
point-like, but perhaps as strings or membranes, similarly to charged
particles e.g., protons and electrons, where the electric charge cannot be
conceived as point-like.   Second, for sufficiently large densities, the four
known interactions are expected to be unified into a superinteraction,
according to supersymmetric theories e.g., [3], where particles could be
``glued'' similarly to quarks into adrons.   In other words, the occurrence of
a superinteraction implies matter returns to a pre-big bang state.

\section{Conclusion}
\label{conc}

Within the context of Newton's theory of gravitation, for an assigned
(point-like) test particle and (point-like) central body, stable circular
orbits in ordinary space have been related to stable circular paths  on a
vortex-like, massless, unmovable, undeformable surface, subjected
to a tidal gravitational field along the symmetry axis, implying constant
attraction directed downwards, and a holographic principle has been proposed.
If ordinary space is conceived as a 3-hypersurface bounding a $n$-hypervolume
where gravitational interaction takes origin, the extension of the above
mentioned
holographic principle has implied a minimum distance behind which test
particles lose theirs individuality (intended as collection of intrinsic
properties)
and ``glue'' to the central body via unification of the four known
interactions, behind which (i) particles can no longer be conceived as
point-like
but perhaps as strings or membranes, and (ii) quantum effects are dominant and
matter returns to a pre-big bang state.

The generalization of the procedure to noncircular motions within the context
of general relativity, together with further knowledge about neutron stars,
quark stars, black holes, e.g., [8] could provide further insight on the
formulation of quantum gravity.


\appendix
\section*{Appendix}

\section{A guidance example}
\label{a:guex}

As a guidance example, let the meridional section of the vortex-like surface
coincide with branches of equilateral hyperbola, as:
\begin{lefteqnarray}
\label{eq:eqhy}
&& wz=\mp k~~;
\end{lefteqnarray}
where the right-hand side is positive or negative according if the hyperbola
branch lies on the fourth or third quadrant, respectively, see
Fig.\,\ref{f:vort}.   Let ${\sf P}\equiv(w_P, z_P)$ be a generic point on the
hyperbola.

The intersection of the hyperbola with a generic straight line, expressed as:
\begin{lefteqnarray}
\label{eq:gere}
&& z=aw+b~~;
\end{lefteqnarray}
can be inferred via substitution of Eq.\,(\ref{eq:gere}) into (\ref{eq:eqhy}),
as solution of the related second-degree equation in $w$.   The result is:
\begin{lefteqnarray}
\label{eq:sols}
&& w_P=\frac{-b\mp\sqrt{b^2\mp4ak}}{2a}~~;
\end{lefteqnarray}
where the double signs are uncorrelated.

The straight line is tangent to an
hyperbola branch provided the discriminant is null, as:
\begin{lefteqnarray}
\label{eq:Del0}
&& b^2\mp4ak=0~~;\qquad\mp ak<0~~;\qquad\pm ak>0~~;
\end{lefteqnarray}
accordingly, the coordinates of the tangential point are:
\begin{lefteqnarray}
\label{eq:wP}
&& w_P=-\frac{b}{2a}=-\frac{\sqrt{\pm ak}}a~~; \\
\label{eq:zP}
&& z_P=\mp\frac k{w_P}=\frac{\pm ak}{\sqrt{\pm ak}}=\sqrt{\pm ak}~~;
\end{lefteqnarray}
from which the slope and the intercepts of the tangent straight line are
inferred via Eqs.\,(\ref{eq:eqhy}) and (\ref{eq:gere}), as:
\begin{lefteqnarray}
\label{eq:a}
&& a=\frac{z_P^2}{\pm k}=-\frac{z_P}{w_P}~~; \\
\label{eq:w0}
&& w_0=-\frac ba=2w_P~~; \\
\label{eq:z0}
&& z_0=b=-2aw_P=2z_P~~;
\end{lefteqnarray}
in terms of the coordinates of the tangential point, ${\sf P}$.   Finally, the
substitution of Eqs.\,(\ref{eq:w0}) and (\ref{eq:z0}) into (\ref{eq:slta})
yields:
\begin{lefteqnarray}
\label{eq:tgtp}
&& \tan\theta=-\frac{z_P}{w_P}~~;
\end{lefteqnarray}
accordingly, $\theta$ is positive or negative if related to an hyperbola branch
lying on the third or the fourth quadrant, respectively, see
Fig.\,\ref{f:vort}.

The substitution of Eqs.\,(\ref{eq:w0}) and (\ref{eq:z0}) into (\ref{eq:wVv})
after little algebra produces:
\begin{lefteqnarray}
\label{eq:wPVv}
&& z_P=\frac{v^2}{V^2}R~~;
\end{lefteqnarray}
which is equivalent to Eq.\,(\ref{eq:zRvV}) and, in particular, $v=V$ implies
$z_P=R$; $v=V=c$ implies $z_P=R_c=GM/c^2$.

\section{An analogy for light velocity invariance}
\label{a:pilv}

Let bodies (particles, in particular) be conceived as spherical vessels moving
on a global ocean.   Let light be conceived as a circular wave on the ocean
which can be induced via e.g., vessel oscillations along
the vertical direction.   Then waves propagate on the ocean surface,
regardless of vessel motion.  The concept of ``light velocity in vacuum''
appears to be meaningless e.g., [9] in that, according to current cosmologies,
space is permeated by a sort of dark energy e.g., [1] and, on the other hand,
quantum vacuum is also permeated by (minimum) energy e.g., [10].

Conceptually, wave velocity can be determined in the following way.
Let vessels move between equally spaced series of buoies, similarly to
electric trains between high-voltage tralisses.   Let $\Delta s$ be the
distance between adjacent buoies (along motion direction), and let $\Delta t$
be the time of wave propagation between adjacent buoies (along motion
direction) which, in principle, can be determined
from vessels.   Accordingly, wave velocity can be inferred as
$v_w=\Delta s/\Delta t$, regardless of vessel motion i.e. reference frame
comoving with a vessel.

In addition, wave velocity cannot be exceeded by vessel velocity, $v<v_w$.
The limit, $v\to v_w$, would imply vessels be turned into water, where waves
can propagate.

Finally, the reference frame comoving with the ocean i.e. at rest with respect
to the ocean can be conceived as ``privileged''.

As mentioned in Section \ref{que2}, ordinary space can be related to ocean
surface and extra dimensions to ocean volume.

\end{document}